\documentclass[11pt,twoside]{article}

\usepackage{asp2014}

\aspSuppressVolSlug
\resetcounters

\begin{document}

\title{JoXSZ -- Joint X-SZ fitter for galaxy clusters}

% Note the position of the comma between the author name and the 
% affiliation number.
% Authors surnames should come after first names or initials, eg John Smith, or J. Smith.
% Author names should be separated by commas.
% The final author should be preceded by "and".
% Affiliations should not be repeated across multiple \affil commands. If several
% authors share an affiliation this should be in a single \affil which can then
% be referenced for several author names. If only one affiliation, no footnotes are needed.
% See ManuscriptInstructions.pdf and ASP's manual2010.pdf 3.1.4 for more details
\author{Fabio~Castagna, Stefano~Andreon}
\affil{INAF--Osservatorio Astronomico di Brera, via Brera 28, 20121 Milan, Italy; \email{fabio.castagna@inaf.it}}

% This section is for ADS Processing.  There must be one line per author. paperauthor has 9 arguments.
\paperauthor{Sample~Author1}{Author1Email@email.edu}{ORCID_Or_Blank}{Author1 Institution}{Author1 Department}{City}{State/Province}{Postal Code}{Country}
\paperauthor{Sample~Author2}{Author2Email@email.edu}{ORCID_Or_Blank}{Author2 Institution}{Author2 Department}{City}{State/Province}{Postal Code}{Country}
\paperauthor{Sample~Author3}{Author3Email@email.edu}{ORCID_Or_Blank}{Author3 Institution}{Author3 Department}{City}{State/Province}{Postal Code}{Country}

% There should be one \aindex line (commented out) for each author. These are used to
% build up the author index for the Proceedings. The surname must come first, followed by
% initials. Note the use of ~ before each initial to control spacing.
% The \author entries (see above) have surname last. These \aindex entries have
% surname first.
% The Aindex.py command willl create them for you after you have constructed the \author
% The first entry should be the first author, for bold-facing the author index page-reference

%\aindex{Castagna,~F.}
%\aindex{Andreon,~S.}

\begin{abstract}
High-resolution observations of the thermal Sunyaev-Zeldovich (SZ) effect and of the X-ray emission of galaxy clusters are becoming more and more widespread, offering us an unique asset to the study of the thermodynamic properties of the intracluster medium.
%allowing us to measure their thermodynamic profiles.
We present \texttt{JoXSZ}, a Bayesian forward-modelling Python code designed to jointly fit the SZ data and the three dimensional X-ray data cube.
%We present \jxsnospace, a bayesian forward modelling Python code that fits the pressure profile of galaxy clusters from SZ data and is optionally embedded in a X-ray data fitter.
%We present \jxsnospace, a bayesian forward modelling Python code to fit the pressure profile of galaxy clusters with the further benefit of performing a joint SZ-plus-X analysis.
%likelihood computation ,,, adopting ... a fast Abel integration, accounting for PSF smearing and transfer function filtering. 
\texttt{JoXSZ} is able to derive the thermodynamic profiles of galaxy clusters for the first time making full and consistent use of all the information contained in the observations. \texttt{JoXSZ} will be
publicly available on GitHub in the near future.
%This is implemented by merging our SZ data processing code \texttt{PreProFit} with \texttt{MBProj2}, a forward modelling X-ray data cube fitter.
%developed by Sanders.

%We present \jxsnospace, a likelihood code to fit the pressure profile of galaxy clusters .... 
%When X-ray data are also available, 
%\jxs likelihood can be 
%allows us to derive the thermodynamic profiles of galaxy clusters
%for the first time making 
%full and consistent use of all the information contained in the observations.
\end{abstract}

% These lines show examples of subject index entries. At this stage these have to commented
% out, and need to be on separate lines. Eventually, they will be automatically uncommented
% and used to generate entries in the Subject Index at the end of the Proceedings volume.
% Don't leave these in! - replace them with ones relevant to your paper.
%\ssindex{FOOBAR!conference!ADASS 2019}
%\ssindex{FOOBAR!organisations!ASP}

% These lines show examples of ASCL index entries. At this stage these have to commented
% out, and need to be on separate lines. Eventually, they will be automatically uncommented
% and used to generate entries in the ASCL Index at the end of the Proceedings volume.
% The ascl.py command will scan your paper on possible code names.
% Don't leave these in! - replace them with ones relevant to your paper.
%\ooindex{FOOBAR, ascl:1101.010}

\section{Introduction}
Galaxy clusters are the largest and most massive gravitationally bound objects in the Universe, and thus they offer a unique tracer of cosmic evolution.
The thermodynamic properties of a galaxy cluster can be gathered from observation in the optical band, the X-ray band or microwaves.
While clusters have been extensively studied using X-ray observations since the end of the 70's, radio measurements via the Sunyaev-Zeldovich (SZ) effect became widespread in the last decade \citep[e.g.,][]{Birkinshaw2005, Mroczkowski2009, Korngut2011, Sayers2013, Adam2015, Romero2017}. 

SZ and X-ray observations both encode information about the intracluster medium.
%Some authors already investigated the potential of a joint analysis of SZ and X-ray data to fit thermodynamic profiles. 
Our aim is to provide the first, to the best of our knowledge, publicly available code for jointly fitting the pressure profile of galaxy clusters.
\texttt{JoXSZ}, as we named it, is built upon the SZ data fitting pipeline described in \texttt{PreProFit} \citep[][source code available on GitHub at \url{https://github.com/fcastagna/preprofit}]{Castagna2019} and an up-dated version of the X-ray data cube fitter \texttt{MBProj2}, originally developed by \citet{Sanders2018}.
A special attention has been given to highly time-consuming operations, since a joint fit is notoriously
slow \citep[e.g.,][]{Ruppin2019}.

%\texttt{JoXSZ} can be a very useful tool for a wide community of astronomers since it is meant to automate %and generalize all the phases of data analyses in an efficient and easy-to-use software pipeline.
%Still, the fit of pressure profile can be extremely slow \citep[e.g.,][]{Ruppin2019} because of the presence of highly time-requiring operations, mostly in the SZ data processing.

\section{Program flow}
As outlined in Fig.~\ref{flow_chart}, the modeling behind \texttt{JoXSZ} relies on the parametrization of three quantities: the pressure profile, described by the generalized Navarro, Frenk \& White (gNFW) model \citep{Nagai2007}, the electronic density profile, represented by a modified $\beta$-model \citep{Vikhlinin2006}, and the metallicity profile, which is assumed to be flat but let free to vary.

\articlefigure{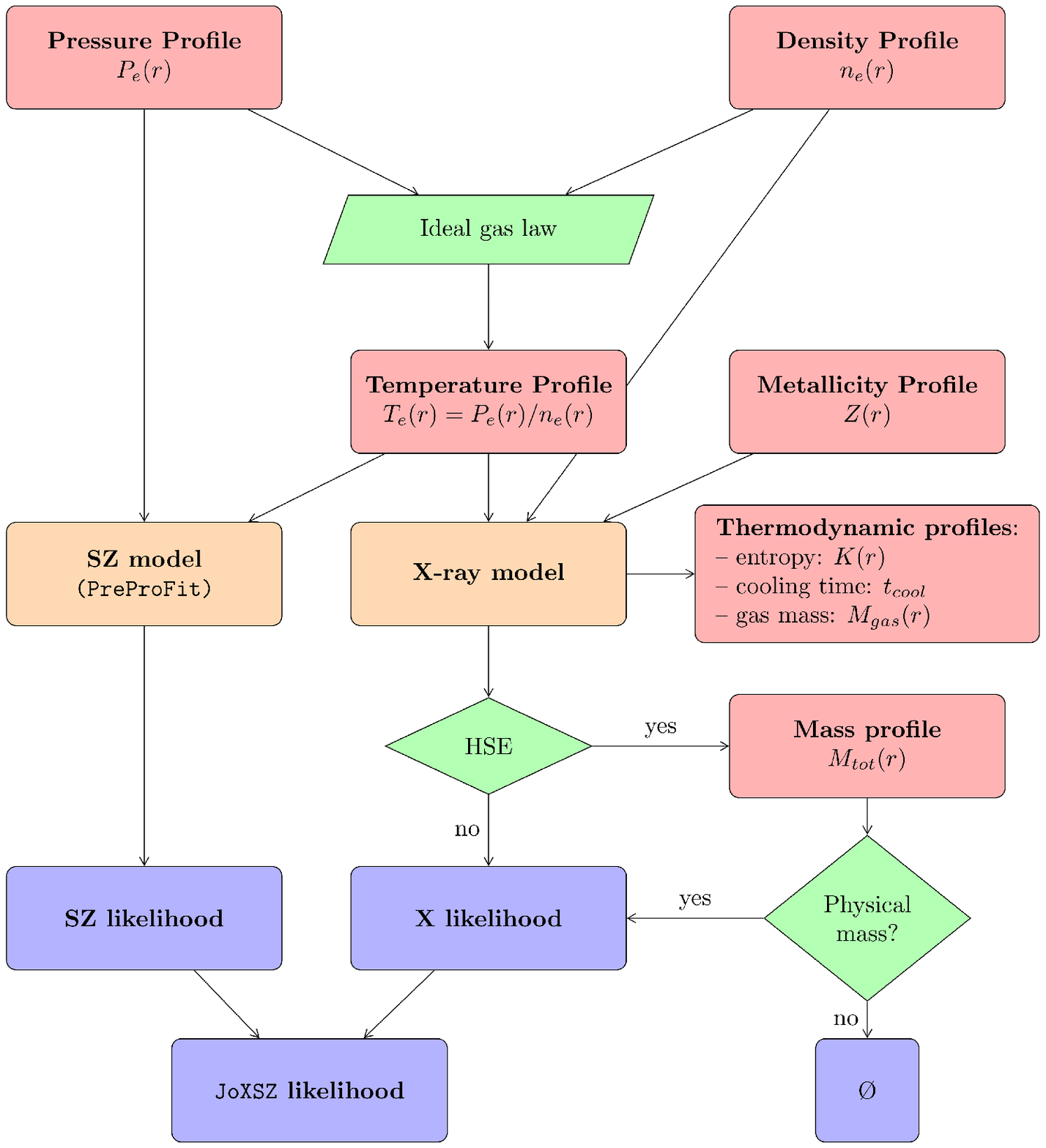}{flow_chart}{Flow chart of \texttt{JoXSZ}. Figure adapted from Castagna \& Andreon in preparation.}
%\articlefiguretwo{P10-9_f1.eps}{P10-9_f2.eps}{flow_chart}{Block diagram showing the program flow.  \emph{Left:} Flow chart of \texttt{JoXSZ}.  \emph{Right:} Flow chart of \texttt{PreProFit}.}
% There is a figure command allowing for three figures:
% \articlefigurethree{P10-9_f1.eps}{P10-9_f1.eps}{P10-9_f1.eps}{ex_fig1_triple}{Now there are three of them.}

The modelization on SZ data, conducted with \texttt{PreProFit}, predominantly involves the pressure profile, though the temperature also takes a part. As outlined in Fig.~\ref{preprofit}, \texttt{PreProFit} projects the three-dimensional pressure profile into a two-dimensional map through the forward Abel transform, then convolves the map with the instrumental beam and the transfer function. Finally, the surface brightness profile is derived through opportune temperature-dependent conversion factors, and the fit to the data is measured by way of the likelihood function of the model.
%Pressure is adopted in the SZ section of the program,
%The SZ pipeline, carried out makes use of the pressure profile

The modelization on X-ray data, conducted with an up-dated version of \texttt{MBProj2}, takes into account the metallicity profile, the density profile and the temperature profile derived as the ratio between pressure and density assuming the ideal gas law.
\texttt{MBProj2} fits surface brightness profiles in multiple X-ray energy bands and automatically computes thermodynamic profiles such as entropy, cooling time and gas mass. Optionally, our up-dated version of \texttt{MBProj2} derives the total mass profile under the assumption of hydrostatic equilibrium adopting a positive prior on mass at all radii. 

\articlefigure{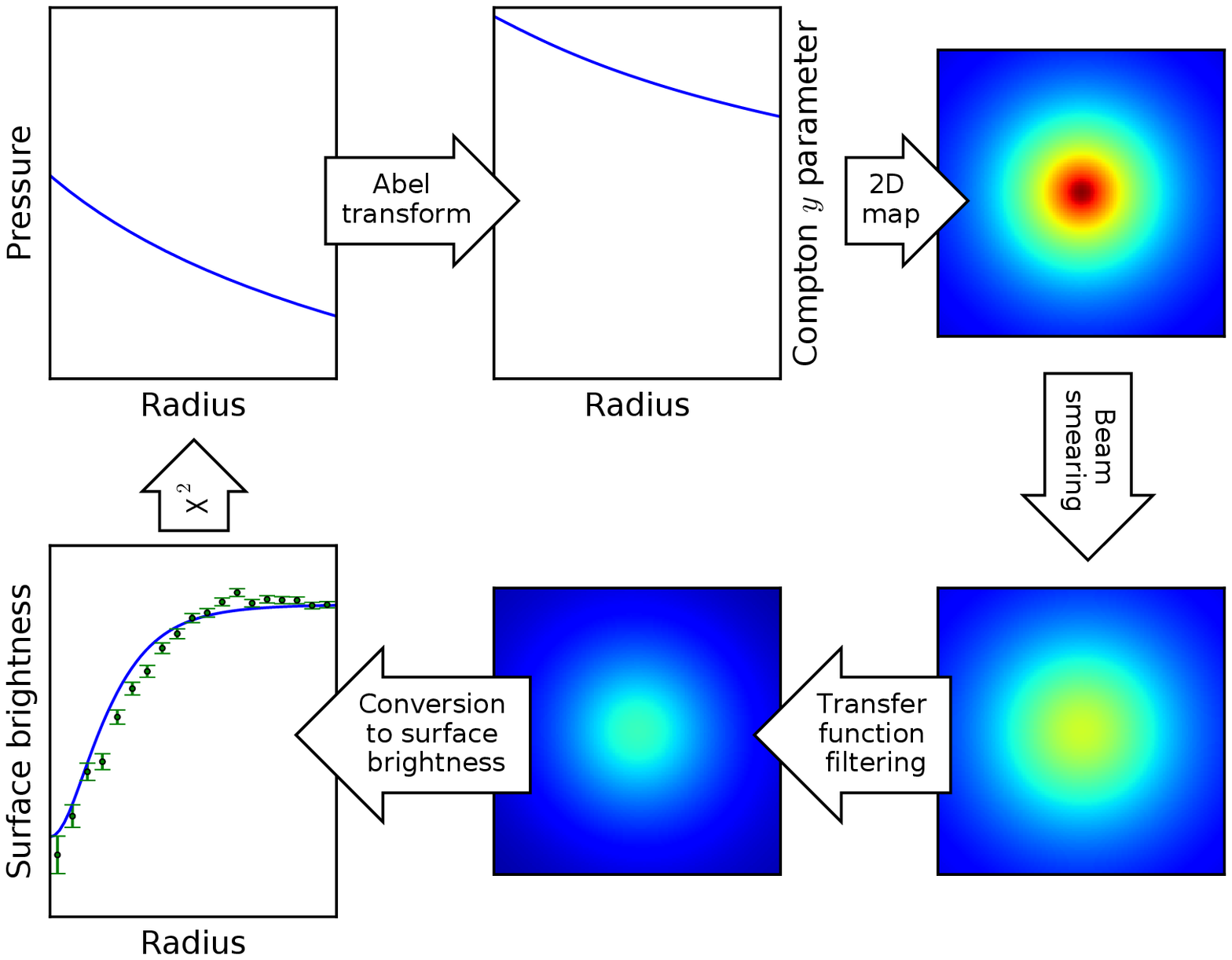}{preprofit}{Block diagram describing the program flow of \texttt{PreProFit}. Figure taken from \citet{Castagna2019}.}
% It is possible to reduce the size of a figure among other changes (see the instructions).  Here is an example:
% \articlefigure[width=.5\textwidth]{P10-9_f1.eps}{ex_fig1_reduced}{Welcome to 1953 a little smaller.}

\section{Functionalities and requirements}

%\texttt{JoXSZ} is able to analyse data coming from different sources and also allows the use of analytic approximations for the beam and transfer functions, useful for feasibility studies, or to make use of published data with only approximate information on these quantities.

\texttt{JoXSZ} relies on a Bayesian forward-modelling approach. The posterior is sampled with \textit{emcee} \citep{Foreman2013} using an affine-invariant ensemble sampler \citep{Goodman2010}.
A large number of parameters are involved, allowing us to fit extremely flexible profiles to the intracluster medium: the gNFW pressure profile has 5 parameters, the Vikhlinin density profile has 7, to which one should add the metallicity $Z$, the temperature ratio $T_{SZ}/T_X$ and a backscale parameter which controls the scaling of the background. As a result, up to 15 parameters can be fitted. Users can specify the list of parameters to fit and is free to select their prior distributions.
The number of random walkers, the number of iterations, the burn-in period extent, and the starting values of the chains can be set by the user as well.

Multi-threading computation is supported by \texttt{JoXSZ} and is strongly encouraged to minimize the time of execution. The SZ model requires the largest amount of execution time (94\%): within the SZ section of the program, Abel transform requires 33\% of the CPU time, two-dimensional image interpolation 18\%, beam smearing 24\%, transfer function filtering 23\%, and other minor operations account for the remaining 2\%.

Qualitative and quantitative diagnostics are both provided by \texttt{JoXSZ} to evaluate the convergence of the chains to the stationary distribution. The acceptance fraction is reported in the program output, as well as the traceplot and the cornerplot are automatically displayed, informing the user of the parameter evolution across iterations and of the joint posterior distribution, respectively. Users can also visualize the surface brightness profile for the best-fitting values and compare it to the observed data.

\texttt{JoXSZ} has been developed and tested with Python 3.6. The following libraries are required to build: \textit{mbproj2, PyAbel, numpy, scipy, astropy, emcee, six, matplotlib, corner}. \texttt{JoXSZ} is planned to be publicly released on GitHub(\url{https://github.com/fcastagna}) at the time of acceptance of the fully referred paper.

\section{Conclusion}
We have presented \texttt{JoXSZ}, the first publicly available code to perform a joint fit of galaxy clusters on both SZ and X-ray data. 
\texttt{JoXSZ} is meant to automate and generalize the whole data analysis process in a consistent and easy-to-use software pipeline. 
It is extensively documented and since it allows the analysis of data coming from different sources, it can be useful to a wide community.
\texttt{JoXSZ} adopts a flexible parametrization of pressure, electron density, temperature and metallicity of the cluster, and efficiently fits these quantities following a Bayesian forward-modelling approach.
Users are free to set up the program in accordance with their needs and requirements: among other things, to decide how many and which parameters to fit, or which prior to adopt for the free parameters.
%, or to conduct a feasibility study using approximations for the beam and the transfer function.

\acknowledgements F.C. acknowleges financial support from the ADASS POC.

\bibliography{P10-9}  % For BibTex

% if we have space left, we might add a conference photograph here. Leave commented for now.
% \bookpartphoto[width=1.0\textwidth]{foobar.eps}{FooBar Photo (Photo: Any Photographer)}

\end{document}